\documentclass{ws-procs961x669}            
\begin{document}
\title{Teaching relativity at the AstroCamp}

\author{C. J. A. P. Martins$^*$}

\address{Centro de Astrof\'{\i}sica da Universidade do Porto, and\\
Instituto de Astrof\'{\i}sica e Ci\^encias do Espa\c co, Universidade do Porto,\\
Rua das Estrelas, 4150-762 Porto, Portugal\\
$^*$E-mail: Carlos.Martins@astro.up.pt}

\begin{abstract}
The AstroCamp is an academic excellence program in the field of astronomy and physics for students in the last 3 years of pre-university education, which often includes a course (or a significant part thereof) on Relativity. After an introduction to the principles, goals and structure of the camp, I describe the approach followed by camp lecturers (myself and others) for teaching Special and General Relativity, and some lessons learned and feedback from the students. I also provide some thoughts on the differences between the physics and mathematics secondary school curricula in Portugal and in other countries, and on how these curricula could be modernized.
\end{abstract}

\keywords{AstroCamp; Secondary education; Summer schools; Relativity.}

\bodymatter

\section{The AstroCamp vision}

The AstroCamp\footnote{\url{https://www.astro.up.pt/astrocamp/}} is an academic excellence program in the field of astronomy and physics, organized by CAUP and the Paredes de Coura municipality (with the support of several national and international partners) for students in the last 3 years of pre-university education, i.e. roughly 15-18 year old students.

Our key goals are
\begin{alphlist}[(c)]
\item[(a)] Promote scientific knowledge, with high-quality training in a secluded and tranquil setting
\item[(b)] Stimulate student curiosity and skills of critical thinking, team work and group responsibility
\item[(c)] Stimulate interactions between students with different backgrounds and life experiences but common interests
\end{alphlist}

The camp was created in 2012, and now accepts applications from 42 eligible countries. Participation is strictly by invitation, after an application period (in April) and a selection phase (in May) which includes an interview in English. The camp itself takes place in August, in the Corno de Bico Protected Landscape area of the Paredes de Coura municipality (in the northwest of Portugal) and lasts 15 days. One of our points of principle is that for students in Portuguese schools the camp has no costs. Some students from other countries will also have support, others may need to pay part of the costs or find their own support locally. In any case, significant efforts are made to keep costs low, and the maximum cost for foreign students is 400 Euro.

The main academic activity consists of two courses (each with 15h of lectures and a 2h written exam), given by currently active researchers with a PhD in a relevant area. Depending on the year, more than two courses are offered, but each student will only take two of them. In most of the 10 editions so far, one course or a significant part thereof was devoted to Relativity (both Special and General). This is to a large extent driven by the camp students themselves: one of the principles of the camp is that students are involved in the choice of the courses offered in the camp, and Relatively is clearly the prime example of a topic that the students feel is not satisfactorily covered in their school classes (if it is at all), and for which their school teachers are often unable to provide further information. 

Other camp activities include observational and/or computational projects, stargazing sessions and documentaries, community service projects and evening talks which in early editions were open to the local public, but are now webcast live. There are also several other recreational activities such as hiking (including an overnight hike coinciding with the peak of the Perseids meteor shower). Interested students can also engage in several post-camp projects at CAUP and elsewhere. Our guiding vision is to provide the students with unique opportunities beyond the standard school systems, also after the camp. The extent to which students make use of all of these is of course up to them.

The camp is residential, with all students, teachers and camp monitors (university students, which most often are AstroCamp alumni) staying in the camp for the full 15 days. (Exceptionally, teachers may stay for reduced periods.) Due to COVID-19, the 2020 edition exceptionally had a hybrid format, with
students living in Portugal in residence at the camp, and students living in
other countries joining virtually in all activities where this was feasible. In the 2021 edition only one student had to participate remotely, and we naturally hope that the 2022 edition will be fully residential again.

\section{Some statistics and outcomes}

In the first four editions the camp only accepted applications from Portuguese students studying in Portuguese schools. Starting in 2016 the list of eligible countries has expanded, and currently the camp can accept applications from students matriculated in one of the last 3 years of pre-university education in any of 42 eligible countries\footnote{These are: Andorra, Argentina, Austria, Belgium, Brazil, Bulgaria, Canada, Chile, Croatia, Cyprus, Czechia, Denmark, Estonia, Finland, France, Germany, Greece, Hungary, Iceland, Ireland, Italy, Latvia, Liechtenstein, Lithuania, Luxembourg, Malta, Monaco, Netherlands, Norway, Poland, Portugal, Romania, San Marino, Serbia, Slovakia, Slovenia, Spain, Sweden, Switzerland, United Kingdom, United States of America and Uruguay.}, provided they are national citizens from one of these countries.

In the 10 editions so far the camp had a total of 133 accepted students with 15 different nationalities. Some basic statistical information on the participants can be found in Table \ref{table1}. It is worthy of note that the camp does not have quotas of any kind; the outcomes are therefore the result of the merit and academic potential of the candidates.
\begin{table}
\tbl{Basic statistical information on the 2012-2021 AstroCamp students.}
{\begin{tabular}{lcc}
\toprule
Accepted Students & 2012-2015 Editions & 2016-2021 Editions\\
\colrule
Portuguese & $100\%$ &  $53\%$ \\
Foreign & N/A &  $47\%$ \\
\colrule
Grade 10 & $24\%$ &  $31\%$ \\
Grade 11 & $52\%$ &  $47\%$ \\
Grade 12 & $24\%$ &  $22\%$ \\
\colrule
Boys & $47\%$ &  $40\%$ \\
Girls & $53\%$ &  $60\%$ \\
\botrule
\end{tabular}}
\begin{tabnote}
Grades 10, 11 and 12 are the names of the last three years of secondary education in the Portuguese school system. The student ages corresponding to these grades are country-specific, but broadly speaking Grade 10 students are 15 or 16 years old, while Grade 12 students are 17 or 18 years old.\\
\end{tabnote}\label{table1}
\end{table}

One of the most rewarding deliverables of the AstroCamp is our Solar System Trail. This is the second scale model of the Solar System, in the Iberian Peninsula (and one of only sixteen in Europe, as far as we know), that is accurate both in terms of the sizes and of the distances of the objects. It was built during the first four AstroCamps, and a full hiking trail which follows the objects in the system---the Sun, the planets up to Neptune (including the Earth's Moon), and Ceres---was officially opened on 13 August 2016. The trail has recently been included among the teaching resources for Grade 7 students, so for a new generation of Portuguese students this may well be providing their first contact with astronomy. The trail also has an accompanying website\footnote{\url{https://www.astro.up.pt/trilhosistemasolar/}} which at the time of writing only has a Portuguese version but is due to be expanded and replicated in at least 10 different languages in early 2022.

Most editions of the camp also included a computational project, lasting about 20 hours, aiming to provide the students with an introduction to scientific programming and statistics and data analysis in astrophysics. In several cases, the students continued working on these projects after the camp, and contributed to peer-reviewed publications in top academic journals \cite{Rocha,Alves,Faria}. At least one more such publication is expected soon.

\section{Teaching Special and General Relativity}

As previously mentioned, One AstroCamp principle is that students are involved in the choices of the courses taught in the camp, as well as in the ones they actually take. This is a multi-stage process, which we run in parallel with the student applications. A call for courses is issued at the beginning of each calendar year, and the submitted proposals undergo a scientific and pedagogical check, leading to a courses shortlist. The students shortlisted  for the interviews are then asked to rank the shortlisted courses according to their preferences; this is done in a double-blind way---students don't know who would teach each course, and organizers don't know individual student preferences until after the student places are assigned. The courses to be offered are then decided and assignment to students, and a final tweaking of course contents can be done if there are overlaps between them. As an example, in the 2021 edition the shortlist included 6 courses, of which 3 were taught in the camp (with each student taking 2 of them).

So far, courses on Relativity/Cosmology (or containing at least ca. 5 hours of content on this) have always been selected by the students when offered. Various lecturers have been involved (in addition to the author), from Canada, Germany, Italy, The Netherlands, Portugal, Spain and the UK, some of them teaching twice.

Topics are covered both at the conceptual and (at least for some of them) also at the more detailed mathematical level. In our experience, most students also enjoy learning about the relevant historical context and background.

Galilean relativity topics covered include
\begin{alphlist}[(e)]
\item[(a)] Galilean mechanics and acceleration
\item[(b)] Inertial frames and Galilean Transformations
\item[(c)] Historical developments from Galileo to Newton (Gassendi, Huygens, Descartes, etc)
\item[(d)] Newton’s Laws and Equivalence Principle (inertial and gravitational masses)
\item[(e)] The XIX Century background to Relativity: Mechanics versus Electromagnetism
\end{alphlist}

Special relativity topics covered include
\begin{alphlist}[(f)]
\item[(a)] Derivation and interpretation of the Lorentz-Fitzgerald transformations
\item[(b)] Minkowski geometry and spacetime diagrams
\item[(c)] Special Relativity and the Principle of Relativity
\item[(d)] Invariance of the speed of light and its consequences
\item[(e)] Time dilation and Lorentz contraction: Cosmic rays and Twin paradox
\item[(f)] Physical meaning of inertial frames
\end{alphlist}

Finally, topics pertaining to applications of relativity covered include
\begin{alphlist}[(e)]
\item[(a)] Relativity and the GPS
\item[(b)] The Schwarzschild solution and black hole properties
\item[(c)] Derivation of the Friedmann and Raychaudhuri equations, and simple solutions thereof
\item[(d)] Cosmology, including Hubble’s law and simple FLRW universes
\item[(e)] Gravitational waves, including their detection
\end{alphlist}

Examples of conceptual questions on these topics that have been previously used in AstroCamp course exams are listed in the Appendix. Student feedback clearly shows that the Newtonian physics curriculum is unattractive, and often containing significant misconceptions. Camp students usually follow conceptual aspects very well (even when finding some of them counter-intuitive); any difficulties usually pertain to their mathematical background, which is partially correlated with their age. 

\section{Lessons learned, and the COVID-19 impact}

The experience of ten camp editions (of which six included foreign students) reveals significant differences between the secondary school curricula in different countries, and sometimes in different regions of the same country. There are also country-specific differences between public and private (or international) schools. In Portugal, it is clear that private schools (and even a few public schools) routinely apply grade inflation. This further reinforces the prior expectation that grades alone do not provide a fair selection process. This is why, from the first edition, the application procedure includes a student motivation letter, a teacher recommendation letter and, importantly, an interview to shortlisted candidates.

Since the camp accepts students from the last three years of secondary education, one challenge for the camp teachers is that the mathematical background will be quite diverse. Specifically, derivatives introduced at different ages in different countries, and integrals are not a given even for Grade 12 students, who have finished secondary education and are about to start university. However, in our experience this is not a problem for bright students: through direct interaction with teachers or camp monitors, they can quickly gain a working knowledge, e.g. of differentiation rules, which allows then the follow the course contents that rely on them.

One rather surprising lesson is that schools can actually be a bottleneck in disseminating information on the AstroCamp to students. Specifically, from our experience less than $20\%$ of Portuguese schools that receive direct camp information from us (including, in some editions, A4 or A3 sized posters and leaflets) actually transmit it to their students, and the fraction of students that find out about the camp through a school teacher is small---while the number is not trivial to estimate, it is clearly below  $50\%$.

A recent challenge was provided by COVID-19. This led to a hybrid format edition in 2020, with foreign students interacting with camp teachers, monitors (including some online tutors) and students in residence at the camp through online collaboration tools such as Zoom, Discord and Colaboratory. online collaboration This was was nevertheless very successful despite the fact that foreign students participating in this edition had a very limited experience of the camp, and clearly demonstrates that remote collaboration tools enable new teaching and mentoring opportunities, which we intend to pursue.

A more surprising realization emerged in the 2021 edition, in that the COVID-19 impact on students’ cognitive and social skills and emotional maturity was clearly visible. Inter alia, although the level of the courses and exams has been kept constant in recent editions (roughly at the level of a course from first year undergraduate physics degree at a top Portuguese university), the average grades of the students were lower and, in particular, for the first time in ten editions two students failed their exams. It is likely that this impact will be noticeable for several years to came. This is all the more scary since these are bright students---the impact on average ones may be much more serious.

As the AstroCamp starts its second decade, this is a good time to revise and update its format, structure and rules, also taking into account the most recent challenges. There is clearly some room for growth---more in extension and scope than in numbers---making the camp an international reference for events of this kind.

\section{Epilogue: Improving Portuguese Secondary Scientific Education}

The experience of 10 years of organizing the AstroCamp (together with teaching one course in each edition, and leading other scientific activities therein) also leads to the conclusion that the Portuguese secondary education system, in scientific areas, is clearly lagging behind those of other developed countries, and needs urgent revision and modernization.

In an effort to contribute to an urgent and much needed debate, I may suggest four very specific measures 

\begin{alphlist}[(d)]
\item[(a)] It is clear that Portuguese students are, on average, far less familiar with scientific programming that their foreign peers. (The Portuguese school system current offers a ‘Computational Applications’ subject, but this is a total waste of time\footnote{As well as an insult to the intelligence of most students, were it not for the fact that it allows an easy high grade to be obtained, which will count towards the students' university application.}.) One must urgently introduce a compulsory scientific programming subject in Grades 10 to 12 grades, using languages such as Python, Octave or Julia, and connected to other subjects, especially Physics, Chemistry, Biology and Mathematics.
\item[(b)] In the Portuguese school system, Physics and Chemistry are taught as a single subject in Grades 10 and 11, and are only separate subject in Grade 12. Moreover, very few students take Grade 12 Physics, either because the school does not offer it or because the students consider it a hard subject (form the perspective of getting a good grade) and prefer softer subjects such as English, Psychology or the infamous Computational Applications. This is a crass mistake. Physics and Chemistry should be separate subjects from Grade 10 (and, arguably, even earlier).
\item[(c)] The Portuguese school system has national exams which, depending on the subject, are either at the end of Grade 11 or Grade 12. The former, which currently include the Physics exam, is a clear mistake. The Physics national exam (and indeed all national exams) should be cumulative and at the end of Grade 12. Moreover, Grade 12 Physics should be made compulsory for students intending to apply for university physics and engineering degrees
\item[(d)] Last but not least, these is a vast disconnect between the secondary school environment and that of university (not to mention that of research, business, etc). A way to bridge the gap would be to reward students with top results in national exams with summer internships at research labs or tech companies (with the student matching done by the hosts). As a way to foster equal opportunities, these internships should not be merely awarded to the students with the numerically highest grades but, for example, to the student with the highest grade in each district (or, more ambitiously, each municipality in Portugal---in the former case this kind of program would reach about 20 students, while in the latter it would reach about 300.
\end{alphlist}

All of these are, at least in principle, easy to implement and would have dramatic and comparatively rapid (as well as quantifiable) impacts. The main bottlenecks, other than political will are our blatant 'teaching to the test approach' (manifest in the blind acceptance of school rankings, but partially stemming from the university numerus clausus system), and the pathological reluctance of teacher unions to any kind of change. The latter can be, at least in part, ascribed to the fact that most secondary school science teachers are in their 50s or 60s. This is especially relevant in the context of the first of the above points, since these teachers and have little computing literacy, and none in coding and scientific programming. Nevertheless, delaying the modernization of the system, with these and other measures, simply increases the personal and financial costs for future generations.

\section*{Acknowledgments}

This work was financed by FEDER---Fundo Europeu de Desenvolvimento Regional funds through the COMPETE 2020---Operational Programme for Competitiveness and Internationalisation (POCI), and by Portuguese funds through FCT - Funda\c c\~ao para a Ci\^encia e a Tecnologia in the framework of the project POCI-01-0145-FEDER-028987 and PTDC/FIS-AST/28987/2017.

\eject

\appendix{Conceptual question examples}

The following is a partial list of relevant conceptual questions that have been used in at least one AstroCamp exam. Except for the last one (with is a proposed essay topic) they are all multiple choice questions.

Two inertial observers measure the speed of the same beam of light and obtain the same finite value. This would surprise Newton because
\begin{alphlist}[(d)]
\item[(a)] According to him the speed of anything should depend on the frame of reference.
\item[(b)] He expected the speed of light to be infinite in inertial frames.
\item[(c)] In inertial reference frames he believed light to be stationary.
\item[(d)] He believed light should have an infinite speed in all frames.
\end{alphlist}

The Michelson-Morley experiments were an attempt to
\begin{alphlist}[(d)]
\item[(a)] Measure the speed of light.
\item[(b)] Test the Lorentz-Fitzgerald transformations.
\item[(c)] Measure the Earth's speed relative to the Sun.
\item[(d)] Detect the Earth's movement relative to the aether.
\end{alphlist}

Relative to an object at rest, a moving object is
\begin{alphlist}[(d)]
\item[(a)] Shorter and younger.
\item[(b)] Longer and older.
\item[(c)] Shorter and older.
\item[(d)] Longer and younger.
\end{alphlist}

Special relativity (SR) and general relativity (GR) corrections must be taken into account for the GPS system to work. If these satellites were moved to LEO,
\begin{alphlist}[(d)]
\item[(a)] Both corrections would decrease.
\item[(b)] The SR correction would decrease and the GR one would increase.
\item[(c)] The SR correction would increase and the GR one would decrease.
\item[(d)] Both corrections would increase.
\end{alphlist}

If an astronaut in a windowless rocket feels pressed against the back of her seat, she won't be able to tell
\begin{alphlist}[(d)]
\item[(a)] Whether the rocket is accelerating or moving at constant speed.
\item[(b)] Whether she is accelerating or resting on the surface of some planet.
\item[(c)] Whether the rocket it moving at constant speed in space or at rest in space.
\item[(d)] Whether she is at rest in space or freely falling towards some planet.
\end{alphlist}

A green animal lives on an extremely dense planet. When observed from space,
\begin{alphlist}[(d)]
\item[(a)] It will look green.
\item[(b)] It will look white.
\item[(c)] It will look blue.
\item[(d)] It will look red.
\end{alphlist}

Which of these is not a direct test of the Einstein Equivalence Principle?
\begin{alphlist}[(d)]
\item[(a)] Measurements of gravitational redshift.
\item[(b)] Tests of the stability of fundamental couplings.
\item[(c)] Measurements of the precession of Mercury's perihelion.
\item[(d)] Tests of the universality of free fall.
\end{alphlist}

Essay topic: Describe the relation between physical reality and the mathematical models with which we describe it. In particular, comment on how this relation may differ in the contexts of the pure and applied physical sciences (e.g., theoretical physics and engineering).

\bibliographystyle{ws-procs961x669}
\bibliography{martinsastrocamp}
\end{document}